\documentclass[pdflatex,sn-mathphys-num]{sn-jnl}

\usepackage{graphicx}%
\usepackage{multirow}%
\usepackage{amsmath,amssymb,amsfonts}%
\usepackage{amsthm}%
\usepackage{mathrsfs}%
\usepackage[title]{appendix}%
\usepackage{xcolor}%
\usepackage{textcomp}%
\usepackage{manyfoot}%
\usepackage{booktabs}%
\usepackage{algorithm}%
\usepackage{algorithmicx}%
\usepackage{algpseudocode}%
\usepackage{listings}%

\usepackage{dcolumn}
\usepackage{bm}
\usepackage{float}
\usepackage{comment}
\usepackage{physics}
\newcommand{\half}{\frac{1}{2}}
\newcommand{\du}{\mathrm{d}u}

\theoremstyle{thmstyleone}%
\theoremstyle{thmstyletwo}%

\theoremstyle{thmstylethree}%

\makeatletter
    \renewcommand{\theequation}{%
    \thesection.\arabic{equation}}
    \@addtoreset{equation}{section}
\makeatother

\begin{document}

\title[Scalar Curvature for the Transverse-Field Ising Model]{Scalar Curvature of the Quantum Exponential Family for the Transverse-Field Ising Model and the Quantum Phase Transition}


\author{\fnm{Takemi} \sur{Nakamura}}\email{nakamura.takemi.d7@s.mail.nagoya-u.ac.jp}

\affil{\orgdiv{Department of Complex Systems Science, Graduate School of Informatics}, \orgname{Nagoya University}, \orgaddress{\city{Nagoya}, \postcode{464-8601}, \country{Japan}}}


\abstract{
Unlike for classical many-body systems, the scalar curvature of the exponential family for quantum many-body systems has been not so investigated, and its physical meaning remains unclear.
In this paper, we analytically study the scalar curvature of the space of Gibbs thermal states, belonging to the quantum exponential family, equipped with the Bogoliubov-Kubo-Mori metric for the zero- and one-dimensional transverse-field Ising model at low and high temperatures.
We find that these scalar curvatures converge to zero in the high-temperature limit whereas they exponentially diverge approaching zero temperature.
This divergence is a consequence of quantumness.
Furthermore, if we can reconsider the criticality of the scalar curvatures at zero temperature, they both can be considered to show a critical behavior with an exponent of 1, and this critical exponent is consistent with the quantum-classical correspondence of the Ising model.
}




\maketitle

\section{Introduction}\label{sec1}

Geometry is useful in physics and is used in various fields such as general relativity\cite{nakahara2018geometry}.
Also in thermodynamics, Riemannian geometry has been applied on the ground that a thermodynamic-state space equips the physically natural Riemannian metric\cite{weinholdMetricGeometryEquilibrium1975,ruppeinerThermodynamicsRiemannianGeometric1979}.
The Riemannian metric is equivalent to the Fisher metric for the Gibbs distribution\cite{crooksMeasuringThermodynamicLength2007}, thus introducing a perspective of information geometry to statistical mechanics\cite{amariMethodsInformationGeometry2000}.
Information geometry views a statistical model as a differentiable manifold, referred to as a statistical manifold, whose coordinate system corresponds to a set of parameters of the model.
In particular, the Gibbs distribution belongs to the exponential family, and control parameters of a thermodynamic system, such as inverse temperature and magnetic field, can be treated as natural parameters of an exponential family.

Ruppeiner\cite{ruppeinerThermodynamicsRiemannianGeometric1979,RuppeinerRiemannianGeometryThermalFluctuation1995} has advanced the Riemannian-geometrical formulation of thermodynamics and has discovered the relationship between the correlation length $\xi$ of a thermodynamic system with a spatial dimension of $d$ and the scalar curvature $R$ of the thermodynamic-state space of the system, which is
\begin{equation}\label{R behavior1}
    R \sim \xi^d ,
\end{equation}
or, using Josephson's scaling law\cite{goldenfeld2018lectures}, equivalently
\begin{equation}\label{R behavior2}
    R \sim \frac{1}{t^{2-\alpha}} ,
\end{equation}
where $\alpha$ stands for the standard critical exponent of the specific heat and $t=|T-T_c|/T_c$ with $T_c$ denoting the critical point of the system.
The relation \eqref{R behavior1} is suggestive: On the one hand, the thermodynamic-state space of the classical ideal gas, which is characterized by the absence of interactions among particles, is flat. On the other hand, the scalar curvature of a thermodynamic system with some interactions is generally nonzero and diverges at the critical point in accordance with the correlation `volume' $\xi^d$.
The relation is validated for several thermodynamic systems\cite{brodyInformationGeometryVapour2008,ruppeinerThermodynamicCurvatureMeasures2010a,ruppeiner2014thermodynamic} (references are therein).
Additionally, the sign of the scalar curvature seems to be physically meaningful, which can be seen distinctively in the analysis of the quantum ideal gas\cite{janyszekRiemannianGeometryStability1990,ruppeiner2014thermodynamic} and the ideal anyon gas\cite{mirza2008ruppeiner,mirza2009nonperturbative}.
These facts suggest that the scalar curvature, a significant quantity in Riemannian geometry, is also significant in thermodynamics and statistical mechanics.

Let us next move on to the Riemannian geometry of a quantum-state space.
If a space consists of pure states, then the physically natural Riemannian metric on the space is considered the Fubini-Study metric\cite{provostRiemannianStructureManifolds1980,woottersStatisticalDistanceHilbert1981}.
For a space of mixed states, however, innumerable Riemannian metrics that are monotone under CPTP maps can be candidates\cite{petzMonotoneMetricsMatrix1996}, in contrast to the space of probability distributions\cite{chentsovStatisticalDecisionRules1982}.
Of special importance among quantum monotone metrics are threefold: the symmetric logarithmic derivative (SLD) or the Bures-Helstrom metric\cite{helstrom1976quantum}, the right logarithmic derivative (RLD) metric\cite{yuen1973multiple}, and the Bogoliubov-Kubo-Mori (BKM) metric\cite{petzGeometryCanonicalCorrelation1994}.
The SLD metric and the RLD metric are useful in quantum parameter estimation\cite{hayashi2016quantum}.
The BKM metric has a close relationship with linear response theory under the name of the canonical correlation in quantum statistical mechanics\cite{kuboStatisticalPhysicsII2012}.
The BKM metric is distinct in that it is the only monotone metric that generally provides the dually flat structure\cite{grasselliUNIQUENESSCHENTSOVMETRIC2001}.
These metrics differ in the way how quantumness appears, and the natural extension of the Fubini-Study metric to a space of mixed states is the SLD metric\cite{Uhlmann1992}.
However, we will adopt the BKM metric here in this work because of the relation with the partition function, which will be discussed later in detail.

We are now particularly interested in the scalar curvature of a quantum-state space.
To the best of our knowledge, Ingarden \textit{et} \textit{al.}\cite{ingarden1982information} firstly have calculated the scalar curvature for a certain quantum system.
They have introduced the BKM metric as a monotone metric to the space of Gibbs states, an example of the quantum exponential family, and have analytically calculated the scalar curvature for several quantum systems; however, they have neither worked on an interacting system nor discussed the physical or statistical meaning of the scalar curvatures they obtained.
Then, we might wonder whether the scalar curvature of a quantum-state space has a physical interpretation.

Nevertheless, the relationship between the scalar curvature of a quantum-state space, particularly the quantum exponential family, and physical properties such as a phase transition is little investigated\cite{zanardiInformationTheoreticDifferentialGeometry2007a,dey2012information}, and what is more, there are few works to calculate the scalar curvature induced from the BKM metric for a specific quantum system\cite{tanakaKuboMoriBogoliubov2006}.
More calculated models may thus be necessary to find the firm physical interpretation of the scalar curvature in the quantum context.

In the present paper, we analytically study the scalar curvature of the quantum exponential family equipped with the BKM metric for the zero-dimensional (0D) and one-dimensional (1D) transverse-field Ising model at low and high temperatures and its criticality along with the quantum phase transition.
Moreover, we introduce to the scalar curvature the perspective of the correspondence between the $(d+1)$ dimensional classical Ising model at finite temperature and the $d$ dimensional transverse field Ising model at zero temperature through the partition function\cite{suzuki1976Relationship}, which provides a reason why here we deal with the 0D Ising model, a toy model without interactions.
From the point of view of the correspondence, we may expect that the critical behavior of the scalar curvature for the transverse-field Ising model is identical to that for the corresponding classical Ising model because these two models should share the same critical exponents, particularly the same $\alpha$.
In other words, from \eqref{R behavior2}, we expect
\begin{equation}\label{R behavior expected}
    R \sim \dfrac{1}{|\Gamma-\Gamma_c|^{2-\alpha}},
\end{equation}
with $\Gamma_c$ denoting a critical transverse field.
However, the SLD metric does not seem to yield this critical behavior\cite{zanardiInformationTheoreticDifferentialGeometry2007a}, where the singularity of the Ising transition is represented by discontinuity.
One of the purposes of this paper is to show that the BKM metric can result in the behavior \eqref{R behavior expected}.

This paper is organized as follows.
In Section \ref{sec2}, we explain the BKM metric in the space of Gibbs states and the geometrical quantities derived from the metric required in the following.
Based on this, in Section \ref{sec3} and Section \ref{sec4}, we deal with the 0D and 1D transverse-field Ising model, respectively.
Lastly, Section \ref{sec5} concludes the paper.

\section{BKM-metric-indued geometry of Gibbs states \label{sec2}}

In this section, we introduce the BKM metric on the space of Gibbs states, which belongs to the quantum exponential family, and the curvature of the metric, with a focus on behavior at low temperature.

Let the Hilbert space of a quantum system of interest be $\mathcal{H}$.
As is well known, a quantum state of the system is generally described by a density operator $\hat{\rho}$, a positive semidefinite operator acting on $\mathcal{H}$ with trace 1.
The set of all density operators on $\mathcal{H}$ is denoted by $\mathcal{S}(\mathcal{H})$:
\begin{equation}
    \mathcal{S}(\mathcal{H}) = \{\hat{\rho} \, | \, \Tr\hat{\rho} = 1, \; \hat{\rho} \ge 0 \} .
\end{equation}

In quantum information geometry, a parametric family of density operators may be viewed as a statistical manifold.
Parameters are generally denoted here by $x=(x^1, x^2,..., x^n)$ with $n$ finite, and the domain of $x$ is a subspace of $\mathbb{R}^n$ denoted by $\mathcal{X}$.
Then, a set of parameterized density operators
\begin{equation}
    \mathcal{M} = \{ \hat{\rho}(x) \; | \;  x \in \mathcal{X} \subset \mathbb{R}^n \} \subset \mathcal{S}(\mathcal{H})
\end{equation}
may be referred to as a quantum statistical manifold.
In the literature, the Gibbs state $e^{-\beta \hat{H}}/Z$ for a Hamiltonian $\hat{H}$ and inverse temperature $\beta = (k_BT)^{-1}$ can reduce to the form of the following quantum exponential family\cite{hasegawaExponentialMixtureFamilies1997}
\begin{equation}\label{quantum exponential family}
    \hat{\rho}(\theta) = \exp[\theta^i \hat{\mathcal{O}}_i - \psi(\theta)] ,
\end{equation}
\begin{equation}
    \psi(\theta) := \ln \Tr[ \exp[\theta^i \hat{\mathcal{O}}_i] ] = \ln Z(\theta).
\end{equation}
Here $\theta = (\theta^1, \theta^2,..., \theta^n) \in \Theta \subset \mathbb{R}^n$ is referred to as the canonical parameter or the natural parameter and can also be regarded as a coordinate system.
$\{ \hat{\mathcal{O}}_i \}_{i=1}^n$ is a set of self-adjoint operators that represent physical observables.
Note that the Einstein summation convention is used throughout this paper.

In the context of statistical mechanics, the potential function $\psi(\theta)$ is given by the logarithm of the partition function $Z(\theta)$ and may be called the Massieu potential or the reduced free energy.
From the viewpoint of the correspondence between the classical Ising model and the transverse-field Ising model through the partition function, the potential functions of these models can be viewed as being equivalent to each other.

We next introduce the BKM metric onto a quantum statistical manifold $\mathcal{M}$ using the quantum relative entropy.
The BKM metric components at $\hat{\rho} \in \mathcal{M} $ in terms of the coordinate system $x$ can be obtained by the Hessian of the quantum relative entropy as\cite{ingarden1982information,hasegawaExponentialMixtureFamilies1997}
\begin{equation}
    g_{ij}(x) = \left. \pdv{}{x^i}{x^j} D(\hat{\rho}(x') || \hat{\rho}(x)) \right|_{x'=x},
\end{equation}
where the quantum relative entropy $D$ is defined as 
\begin{equation}
    D(\hat{\rho} || \hat{\sigma}) := \Tr[\hat{\rho} (\ln \hat{\rho}-\ln \hat{\sigma}) ] .
\end{equation}
Particularly for a density operator of the exponential form
\begin{equation}
    \hat{\rho}(x) = \exp[\hat{A}(x) - \psi(x)] ,
\end{equation}
the BKM metric components can be reduced to the following simple expression
\begin{equation}\label{BKM metric}
    g_{ij}(x) = \frac{\partial^2 \psi(x)}{\partial x^i \partial x^j} - \Tr[\hat{\rho}(x) \frac{\partial^2 \hat{A}(x)}{\partial x^i \partial x^j}] .
\end{equation}
Moreover, for the quantum exponential family \eqref{quantum exponential family} in terms of the canonical parameterization $\theta$, the BKM metric components can be obtained by the Hessian of the potential function:
\begin{equation}\label{BKM metric for canonical parameter}
    g_{ij}(\theta) = \partial_i \partial_j \psi(\theta) = \int_0^1 \du \Tr[\dfrac{1}{Z} e^{(1-u) \theta^k \mathcal{\hat{O}}_k} (\mathcal{\hat{O}}_i - \expval*{\mathcal{\hat{O}}_i}) e^{u \theta^k \mathcal{\hat{O}}_k} (\mathcal{\hat{O}}_j - \expval*{\mathcal{\hat{O}}_j})],
\end{equation}
where $\partial_i$ denotes $\partial/\partial\theta^i$, the derivative with respect to the canonical parameter $\theta^i$.
This means that canonical parameterization makes the calculation of the BKM metric simple.
This Hessian structure is exactly the same as in the classical case, where the Fisher metric is also given by the Hessian of the potential function of the exponential family.
Thus, from the perspective of the quantum-classical correspondence through the partition function (recall that the potential function is given simply by the logarithm of the partition function), adopting the BKM metric as a Riemannian metric on the quantum exponential family is seemingly reasonable.
Indeed, in the context of statistical mechanics, \eqref{BKM metric for canonical parameter} gives a susceptibility and is likely to show a critical behavior, whose critical exponent is, according to the correspondence, common between the two corresponding quantum and classical models.

When a Riemannian metric is a Hessian metric, as given in the equation \eqref{BKM metric for canonical parameter}, the Christoffel symbols $\Gamma_{ijk}$ and the Riemannian curvature tensor $R_{ijkl}$ take rather simple forms \cite{janyszekRiemannianGeometryThermodynamics1989}:
\begin{align}
    \Gamma_{ij;\,k}(\theta) &:= \half (\partial_i g_{jk} + \partial_jg_{ki} - \partial_kg_{ij}) = \half \psi_{ijk}(\theta) , \label{Christoffel symbol for Gibbs state} \\
    R_{ijkl}(\theta) &:= \dfrac{1}{2} \qty(\partial_j\partial_l g_{ik} - \partial_j\partial_k g_{il} + \partial_i\partial_k g_{jl} - \partial_i\partial_l g_{jk}) + g^{ab} (\Gamma_{jl;a} \Gamma_{ik;b} - \Gamma_{jk;a} \Gamma_{il;b}) \notag \\
    &= \frac{1}{4} g^{ab} (\psi_{aik} \psi_{bjl} - \psi_{ail} \psi_{bjk} ) \label{def of scalar curvature},
\end{align}
where $\partial_i \partial_j \partial_k \psi(\theta)$ is abbreviated as $\psi_{ijk}(\theta)$.
Here we define the Riemannian curvature tensor as the curvature of the sphere becomes negative, in accordance with Ruppeiner\cite{RuppeinerRiemannianGeometryThermalFluctuation1995,ruppeinerThermodynamicCurvatureMeasures2010a,ruppeiner2014thermodynamic}.
Note that this sign convention is opposite to the one more frequently used.
Due to the Hessian structure, the Christoffel symbols can be obtained just by the third derivative of the potential function, and the Riemannian curvature tensor does not have fourth derivatives because they cancel out with each other.

For the two-dimensional quantum exponential family with two canonical parameters $\theta = (\theta^1, \theta^2)$, which is our main concern, the scalar curvature can be calculated by this useful expression\cite{janyszekRiemannianGeometryThermodynamics1989}
\begin{equation}\label{scalar curvature in 2-dim for Gibbs state}
    R(\theta^1,\theta^2) 
    = \dfrac{2 R_{1212}}{\det g}
    = \dfrac{\mdet{g_{11} & g_{12} & g_{22} \\ \psi_{111} & \psi_{112} & \psi_{122} \\ \psi_{112} & \psi_{122} & \psi_{222} }}{2 \mdet{g_{11} & g_{12} \\ g_{21} & g_{22}}^2} .
\end{equation}
This expression results essentially from a Hessian metric, and thus can be used both for quantum and classical systems as long as a Riemannian metric is Hessian.
In fact, it has been often used for calculating the scalar curvature of a thermodynamic-state space\cite{brodyInformationGeometryVapour2008,ruppeinerThermodynamicCurvatureMeasures2010a,ruppeiner2014thermodynamic}.
For the discussion below, let $F$ denote the numerator:
\begin{equation}\label{numerator}
    F := \mdet{g_{11} & g_{12} & g_{22} \\ \psi_{111} & \psi_{112} & \psi_{122} \\ \psi_{112} & \psi_{122} & \psi_{222} }.
\end{equation}

In the quantum-classical correspondence, the model on the quantum side requires zero temperature.
To cope with zero temperature, we adjust the dimension and "normalize" the potential with the inverse temperature $\beta$, that is, we introduce the negative free energy $1/\beta \ln Z$ as the potential instead of the Massieu potential $\ln Z$.
In this case, the BKM metric will be normalized accordingly and may be redefined as the canonical correlation
\begin{equation}\label{scaled BKM metric}
    g_{ij}(\theta) = \dfrac{1}{\beta} \partial_i \partial_j \ln Z(\theta) = \dfrac{1}{\beta} \int_0^1 \du \Tr[\dfrac{1}{Z} e^{(1-u) \theta^k \mathcal{\hat{O}}_k} \mathcal{\hat{O}}_i e^{u \theta^k \mathcal{\hat{O}}_k} \mathcal{\hat{O}}_j] - \expval*{\mathcal{\hat{O}}_i}\expval*{\mathcal{\hat{O}}_j}.
\end{equation}
For this new metric, the scalar curvature has the same dimension as $\beta$, i.e., $(\text{energy})^{-1}$.

To discuss low-temperature behavior, we consider the representation by the energy basis $\{\ket{n}\}$ that diagonalizes a Hamiltonian $\hat{H}$: $\hat{H} \ket{n} = E_n \ket{n}$, or with $\beta$ included,
\begin{equation}\label{eigenvalue equation}
    -\beta \hat{H} \ket{n} = \theta^i \hat{\mathcal{O}} \ket{n} = \mathcal{E}_n \ket{n},
\end{equation}
where $\mathcal{E}_n := -\beta E_n$.
In the following, we assume that $\hat{H}$ is non-degenerate for simplicity.
In this representation, the BKM metric components \eqref{scaled BKM metric} can be written concretely as
\begin{equation}\label{BKM metric for canonical parameter: energy basis}
    g_{ij}(\theta) = g_{ij}^{\mathrm{c}}(\theta) + g_{ij}^{\mathrm{q}}(\theta),
\end{equation}
where
\begin{equation}\label{BKM metric: classical part}
    \beta g_{ij}^{\mathrm{c}}(\theta) := \sum_n p_n^{\mathrm{G}} (\mathcal{\hat{O}}_i)_{nn}  (\mathcal{\hat{O}}_j)_{nn} - \expval*{\mathcal{\hat{O}}_i} \expval*{\mathcal{\hat{O}}_j},
\end{equation}
\begin{equation}\label{BKM metric: quantum part}
    \beta g_{ij}^{\mathrm{q}}(\theta) := - \dfrac{1}{\beta} \sum_{n\neq m} \dfrac{p_m^{\mathrm{G}}-p_n^{\mathrm{G}}}{E_m - E_n} (\mathcal{\hat{O}}_i)_{nm}  (\mathcal{\hat{O}}_j)_{mn},
\end{equation}
and $(\mathcal{\hat{O}}_i)_{nm} := \mel{n}{\mathcal{\hat{O}}_i}{m}$ is a matrix element of $\mathcal{\hat{O}}_i$ in the energy basis, $p_n^{\mathrm{G}} = e^{-\beta E_n}/Z$ represents the Gibbs distribution for the energy $E_n$, and $\expval*{\mathcal{\hat{O}}_i}$ is an expectation value with respect to the Gibbs distribution defined by $\expval*{\mathcal{\hat{O}}_i} := \sum_n p_n^{\mathrm{G}} (\mathcal{\hat{O}}_i)_{nn}.$
As shown in \eqref{BKM metric for canonical parameter: energy basis}, the BKM metric can be separated into two parts.
The first term $g_{ij}^{\mathrm{c}}(\theta)$ can be viewed as thermal fluctuation in that it consists only of diagonal elements and is equal to the classical Fisher information $\mathbb{E}_p[-\partial_i \partial_j \ln p]$ with respect to the Gibbs distribution $\{p_n^{\mathrm{G}}\}_n$, whereas the second one $g_{ij}^{\mathrm{q}}(\theta)$ as quantum fluctuation in that it involves non-diagonal elements and reflects non-commutativity among observables.
It should be noted that from this expression, one can immediately check that the quantum part in the BKM metric is different from that in the SLD metric, which is given specifically e.g. in \cite{hubner1992explicit, zanardi2007bures}.
Therefore, it is natural that curvature properties should be different between the two metrics as long as quantumness is considered.

As expected, the thermal fluctuation $\beta g_{ij}^{\mathrm{c}}(\theta)$ vanishes at zero temperature:
\begin{equation}
    \lim_{T\to0} \beta g_{ij}^{\mathrm{c}}(\theta) = 0,
\end{equation}
but quantum fluctuation $\beta^2 g_{ij}^{\mathrm{q}}(\theta)$ generally remains nonzero and given as
\begin{equation}
    \lim_{T\to0} \beta^2 g_{ij}^{\mathrm{q}}(\theta) = \sum_{n>0} \dfrac{1}{E_n - E_0} 2\Re (\mathcal{\hat{O}}_i)_{n0} (\mathcal{\hat{O}}_j)_{0n} =: g_{(0)ij}^{\mathrm{q}}(\theta),
\end{equation}
and it can have a singularity at a quantum phase transition point.
However, from the constraint $\theta^i (\mathcal{\hat{O}}_i)_{0n} = 0$ for $n > 0$, which can be obtained from the eigenvalue equation \eqref{eigenvalue equation}, we can directly confirm that its determinant vanishes at zero temperature:
\begin{equation}
    \lim_{T\to0} \det(\beta^2 g_{ij}^{\mathrm{q}}(\theta)) = 0.
\end{equation}
This means that the BKM metric becomes degenerate at zero temperature even if there is a contribution from quantumness.
We also find that $F \to 0$ as $T\to 0$ because of the constraint
\begin{equation}\label{constraint: psi at T=0}
    (\theta_1 \partial_1 + \theta_2 \partial_2) g_{(0)ij}^{\mathrm{q}} + g_{(0)ij}^{\mathrm{q}} = 0,
\end{equation}
which can be obtained by the relations $\partial_i \mathcal{E}_n = (\mathcal{\hat{O}}_i)_{nn}$ and $(\mathcal{E}_n-\mathcal{E}_m) \braket{m}{\partial_i n} = (\mathcal{\hat{O}}_i)_{mn}$ following from the eigenvalue equation \eqref{eigenvalue equation}.
Therefore, the scalar curvature \eqref{scalar curvature in 2-dim for Gibbs state} is indefinite as long as we care for the metric only at zero temperature.

Then, to determine the scalar curvature at zero temperature from the formula \eqref{scalar curvature in 2-dim for Gibbs state}, we need to consider at least up to the first-order term.
Here we deal with the situation where $\beta$ is large compared to the minimum energy gap $\Delta := E_1 -E_0$, and contributions from excited states decay exponentially as $T\to 0$, such as disordered phases, so that we can expand the BKM metric up to the first order as
\begin{equation}
    g_{ij}^{\mathrm{c}}(\theta) \simeq 0 + \dfrac{1}{\beta} g_{(1)ij}^{\mathrm{c}}(\theta) \varepsilon,
\end{equation}
\begin{equation}
    g_{ij}^{\mathrm{q}}(\theta) \simeq \dfrac{1}{\beta^2} g_{(0)ij}^{\mathrm{q}}(\theta) + \dfrac{1}{\beta^2} g_{(1)ij}^{\mathrm{q}}(\theta) \varepsilon,
\end{equation}
where $\varepsilon := e^{-\beta \Delta}$.
As for $g_{(1)ij}^{\mathrm{c}}$, one can obtain
\begin{equation}\label{BKM metric: first order classical part}
    g_{(1)ij}^{\mathrm{c}} = (\partial_i \beta \Delta) (\partial_j \beta \Delta) = \qty((\mathcal{\hat{O}}_i)_{00} - (\mathcal{\hat{O}}_i)_{11}) \qty((\mathcal{\hat{O}}_j)_{00} - (\mathcal{\hat{O}}_j)_{11}).
\end{equation}
Hence, the determinant of the BKM metric for a two-dimensional quantum exponential family is calculated, up to the first order of $\varepsilon$, as
\begin{equation}
    \dfrac{\det(g_{ij})}{\varepsilon} \simeq %
    \mdet{\dfrac{1}{\beta} g_{(1)11}^{\mathrm{c}} + \dfrac{1}{\beta^2} g_{(1)11}^{\mathrm{q}} & \dfrac{1}{\beta^2} g_{(0)12}^{\mathrm{q}} \\ %
    \dfrac{1}{\beta} g_{(1)21}^{\mathrm{c}} + \dfrac{1}{\beta^2} g_{(1)21}^{\mathrm{q}} & \dfrac{1}{\beta^2} g_{(0)22}^{\mathrm{q}}} %
    + \mdet{\dfrac{1}{\beta^2} g_{(0)11}^{\mathrm{q}} & \dfrac{1}{\beta} g_{(1)12}^{\mathrm{c}} + \dfrac{1}{\beta^2} g_{(1)12}^{\mathrm{q}} \\ %
    \dfrac{1}{\beta^2} g_{(0)21}^{\mathrm{q}} & \dfrac{1}{\beta} g_{(1)22}^{\mathrm{c}} + \dfrac{1}{\beta^2} g_{(1)22}^{\mathrm{q}} }.
\end{equation}
If we can expect $g_{(1)ij}^{\mathrm{c}} \gg \frac{1}{\beta}g_{(1)ij}^{\mathrm{q}}$ at sufficiently low temperature $\beta \Delta \gg 1$, the leading term of the right-hand side of this is
\begin{equation}\label{denominator: first order coefficient}
    \dfrac{\det(g_{ij})}{\varepsilon} \simeq %
    \dfrac{1}{\beta^3} \qty( \mdet{g_{(1)11}^{\mathrm{c}} & g_{(0)12}^{\mathrm{q}} \\ %
    g_{(1)21}^{\mathrm{c}} & g_{(0)22}^{\mathrm{q}}} %
    + \mdet{g_{(0)11}^{\mathrm{q}} & g_{(1)12}^{\mathrm{c}} \\ %
    g_{(0)21}^{\mathrm{q}} & g_{(1)22}^{\mathrm{c}} } ).
\end{equation}
A similar reasoning yields (see Appendix \ref{secA1})
\begin{equation}\label{numerator: first order coefficient}
    \frac{F}{\varepsilon} 
    \simeq \dfrac{1}{\beta^6} \qty( %
    \mdet{g_{(0)11}^{\mathrm{q}} & g_{(0)12}^{\mathrm{q}} & g_{(0)22}^{\mathrm{q}} \\ %
    \psi_{(1)111}^{\mathrm{c}} & \psi_{(1)112}^{\mathrm{c}} & \psi_{(1)122}^{\mathrm{c}} \\ %
    \psi_{(0)112}^{\mathrm{q}} & \psi_{(0)122}^{\mathrm{q}} & \psi_{(0)222}^{\mathrm{q}} } %
    + \mdet{g_{(0)11}^{\mathrm{q}} & g_{(0)12}^{\mathrm{q}} & g_{(0)22}^{\mathrm{q}} \\ %
    \psi_{(0)111}^{\mathrm{q}} & \psi_{(0)112}^{\mathrm{q}} & \psi_{(0)122}^{\mathrm{q}} \\ %
    \psi_{(1)112}^{\mathrm{c}} & \psi_{(1)122}^{\mathrm{c}} & \psi_{(1)222}^{\mathrm{c}} } %
    ),
\end{equation}
where $\psi_{(0)ijk}^{\mathrm{q}} = \partial_k g_{(0)ij}^{\mathrm{q}}$ and
\begin{equation}
    \psi_{(1)ijk}^{\mathrm{c}} = \qty((\mathcal{\hat{O}}_i)_{00} - (\mathcal{\hat{O}}_i)_{11}) \qty((\mathcal{\hat{O}}_j)_{00} - (\mathcal{\hat{O}}_j)_{11}) \qty((\mathcal{\hat{O}}_k)_{00} - (\mathcal{\hat{O}}_k)_{11}).
\end{equation}
Therefore, unless $F/\varepsilon = 0$ identically, the asymptotic behavior of the scalar curvature \eqref{scalar curvature in 2-dim for Gibbs state} for $T\to 0$ can be obtained as
\begin{equation}\label{R behavior at T=0}
    R(\theta^1, \theta^2) \simeq C e^{\beta \Delta},
\end{equation}
where the coefficient $C$ is defined as
\begin{equation}\label{C}
    C := \dfrac{\mdet{g_{(0)11}^{\mathrm{q}} & g_{(0)12}^{\mathrm{q}} & g_{(0)22}^{\mathrm{q}} \\ %
    \psi_{(1)111}^{\mathrm{c}} & \psi_{(1)112}^{\mathrm{c}} & \psi_{(1)122}^{\mathrm{c}} \\ %
    \psi_{(0)112}^{\mathrm{q}} & \psi_{(0)122}^{\mathrm{q}} & \psi_{(0)222}^{\mathrm{q}} } %
    + \mdet{g_{(0)11}^{\mathrm{q}} & g_{(0)12}^{\mathrm{q}} & g_{(0)22}^{\mathrm{q}} \\ %
    \psi_{(0)111}^{\mathrm{q}} & \psi_{(0)112}^{\mathrm{q}} & \psi_{(0)122}^{\mathrm{q}} \\ %
    \psi_{(1)112}^{\mathrm{c}} & \psi_{(1)122}^{\mathrm{c}} & \psi_{(1)222}^{\mathrm{c}} }}{2 \qty( \mdet{g_{(1)11}^{\mathrm{c}} & g_{(0)12}^{\mathrm{q}} \\ %
    g_{(1)21}^{\mathrm{c}} & g_{(0)22}^{\mathrm{q}}} %
    + \mdet{g_{(0)11}^{\mathrm{q}} & g_{(1)12}^{\mathrm{c}} \\ %
    g_{(0)21}^{\mathrm{q}} & g_{(1)22}^{\mathrm{c}} } )^2}.
\end{equation}
It should be noted that this coefficient has the dimension of $(\text{energy})^{-1}$ and does not depend on $\beta$ at all.

This exponential divergence of the scalar curvature as $T\to 0$, or the form \eqref{R behavior at T=0}, will not arise in the classical case in general because, as $T\to 0$, $F=\order{\varepsilon^4}$ and $\det(g_{ij}^{\mathrm{c}}) = \order{\varepsilon^2}$, yielding $R = \order{\varepsilon^4}/\order{\varepsilon^4} = \order{\varepsilon^0}$.
The behavior of $F$ is due to
\begin{align}
    \mdet{g_{(1)11}^{\mathrm{c}} & g_{(1)12}^{\mathrm{c}} & g_{(1)22}^{\mathrm{c}} \\ %
    \psi_{(1)111}^{\mathrm{c}} & \psi_{(1)112}^{\mathrm{c}} & \psi_{(1)122}^{\mathrm{c}} \\ %
    \psi_{(1)112}^{\mathrm{c}} & \psi_{(1)122}^{\mathrm{c}} & \psi_{(1)222}^{\mathrm{c}} } 
    =& \qty((\mathcal{\hat{O}}_1)_{00} - (\mathcal{\hat{O}}_1)_{11})^4 \qty((\mathcal{\hat{O}}_2)_{00} - (\mathcal{\hat{O}}_2)_{11})^4 %
    \mdet{1 & 1 & 1 \\ %
    1 & 1 & 1 \\ %
    1 & 1 & 1 } \notag \\
    =& 0.
\end{align}
In other words, the divergence in general is a result fully of quantumness since it is attributed to the existence of the quantum contribution in the BKM metric \eqref{BKM metric: quantum part}.

The coefficient $C$ can be considered to possess information about a continuous quantum phase transition since it involves quantum susceptibilities $g_{(0)ij}^{\mathrm{q}}$ and their derivatives $\psi_{(0)ijk}^{\mathrm{q}} = \partial_k g_{(0)ij}^{\mathrm{q}}$.
In other words, $C$ can have a critical behavior according to $\Delta \to 0$.
But the problem here is that \eqref{R behavior at T=0} says the scalar curvature diverges exponentially in $T\to 0$ because of the factor $e^{\beta \Delta}$, in contrast to the susceptibility $g_{(0)ij}^{\mathrm{q}}$, where critical exponents like $\alpha$ are defined without such a problem.
This situation prevents us from taking the limit $\Delta \to 0$ for the scalar curvature \eqref{R behavior at T=0} as it is.
However, if one can regard the criticality of the scalar curvature as $C$ itself, then one can define the critical exponent of the scalar curvature even in this situation.
The exponential divergence itself appears regardless of whether a system has a phase transition or long-range order, and hence it would be reasonable to consider them separately when we discuss a quantum phase transition at zero temperature.
We will later see the quantum-classical correspondence hold for the scalar curvature if we admit and adopt this definition.

In the following two sections, we will confirm the behavior \eqref{R behavior at T=0} and the critical behavior of $C$ considering the transverse-field Ising models.

\section{The zero-dimensional transverse-field Ising model \label{sec3}}

Riemannian-geometric quantities concerning a single spin 1/2 (or qubit) have already been calculated\cite{ingarden1982information, balian2014entropy}.
The 0D transverse-field Ising model, a special case of that, is defined by the Hamiltonian
\begin{equation}
    \hat{H}_{0\mathrm{D}} = -\Gamma \hat{\sigma}^x - h \hat{\sigma}^z,
\end{equation}
or with $x:=\beta \Gamma$ and $z:=\beta h$,
\begin{equation}\label{zero-dimensional transverse-field Ising model}
    -\beta \hat{H}_{0\mathrm{D}} = x\hat{\sigma^x} + z\hat{\sigma^z} ,
\end{equation}
where $\beta=(k_BT)^{-1}$ denotes the inverse temperature, and $\hat{\sigma^x}$ and $\hat{\sigma^z}$ are the Pauli matrices
\begin{equation}
    \hat{\sigma^x} = \mqty(0 & 1 \\ 1 & 0) , \qquad \hat{\sigma^z} = \mqty(1 & 0 \\ 0 & -1) .
\end{equation}
In this case, the parameters $x,z \in (0,\infty)$ are the canonical parameters
\footnote{Using the letters $x,z$ instead of $\theta$ reflects the Bloch sphere representation.}.
Note that this model has two energy eigenvalues $\pm \sqrt{h^2+\Gamma^2}$ and the energy gap for this model is thus given by $2\sqrt{h^2+\Gamma^2}$.
One can justify the use of statistical mechanics for this model by considering the $N\gg 1$ independent spins and physical quantities per spin.

The potential or the negative free energy for this model is easily calculated to become
\begin{equation}
    \psi_{0\mathrm{D}}(\theta, x) := \dfrac{1}{\beta} \ln Z_{0\mathrm{D}}(x,z) = \dfrac{1}{\beta} \ln 2\cosh r 
\end{equation}
with the polar coordinate
\begin{equation}
    r:=\sqrt{z^2+x^2} .
\end{equation}
We thus obtain the metric and the Christoffel symbols as
\begin{gather}
    \beta \qty(g_{0\mathrm{D}})_{xx} = z^2 r^{-3} \tanh r + x^2r^{-2} \cosh^{-2}r , \\
    \beta \qty(g_{0\mathrm{D}})_{xz} = -xzr^{-3}\qty(\tanh r - r\cosh^{-2}r) , \\
    \beta \qty(g_{0\mathrm{D}})_{zz} = x^2r^{-3}\tanh r + z^2r^{-2}\cosh^{-2}r , \\
    \beta^2 \det\qty(g_{0\mathrm{D}}) = r^{-1}\tanh r \cosh^{-2}r , \label{det of BKM metric of 0D Ising}
\end{gather}
\begin{align}
    \beta (\psi_{0\mathrm{D}})_{xxx} &= 
    -3z^2xr^{-5}\tanh r + xr^{-4} (3z^2-2x^2r\tanh r)\cosh^{-2}r , \\
    \beta (\psi_{0\mathrm{D}})_{xxz} &= 
    -zr^{-5} (r^2-3x^2)\tanh r + zr^{-4} (r^2-3x^2-2x^2r\tanh r) \cosh^{-2}r , \\
    \beta (\psi_{0\mathrm{D}})_{xzz} &= 
    -xr^{-5} (r^2-3z^2)\tanh r + xr^{-4} (r^2-3z^2-2z^2r\tanh r) \cosh^{-2}r , \\
    \beta (\psi_{0\mathrm{D}})_{zzz} &= 
    -3x^2zr^{-5}\tanh r + zr^{-4} (3x^2-2z^2r\tanh r)\cosh^{-2}r .
\end{align}
Therefore, the Riemannian curvature tensor and the scalar curvature are calculated from the equation \eqref{scalar curvature in 2-dim for Gibbs state} as 
\begin{gather}
    \qty(R_{0\mathrm{D}})_{xzxz} = \dfrac{1}{\beta} \qty(\dfrac{2r-\tanh r}{4r^3} - \dfrac{1+\tanh^2r}{4r\tanh r\cosh^2r}) ,
\end{gather}
\begin{equation}\label{scalar curvature of 0D quantum Ising}
    R_{0\mathrm{D}} = \beta \qty(\dfrac{2r-\tanh r}{2r^2\tanh r}\cosh^2r - \dfrac{1+\tanh^2r}{2\tanh^2r}) .
\end{equation}

We get a physically notable result here.
Although the 0D transverse-field Ising model has no physical interaction, the scalar curvature for this model does not vanish.
This may be because the quantum effect, more precisely the non-commutativity of the Pauli matrices, causes effective correlations, as in the case of the quantum ideal gas\cite{janyszekRiemannianGeometryStability1990,pessoaInformationGeometryFermiDirac2021}.
Indeed, if we can ignore the contribution of the non-commutativity between $\hat{\sigma}_x$ and $\hat{\sigma}_z$ so that the exponential can be decomposed classically as
\begin{equation}
    e^{x\hat{\sigma}_x + z\hat{\sigma}_z} \simeq e^{x\hat{\sigma}_x} e^{z\hat{\sigma}_z},
\end{equation}
then the potential becomes
\begin{equation}
     \ln Z_{0\mathrm{D}}(x,z) \simeq \ln2 + \dfrac{x^2+z^2}{2},
\end{equation}
and we observe that the metric becomes Euclidean, resulting in flatness.
The condition for this classicality is $r \ll 1$, which means that the system is either at sufficiently high temperature or in the vicinity of the critical point $r=0$ with $\beta$ finite.
Behavior in this region is revisited later.
This classical approximation breaks down if we cannot ignore the quantumness and need to consider higher terms for the potential.

The same situation occurs in the discussion of the quantum-classical correspondence of the Ising model\cite{suzuki1976Relationship}. 
When we map the transverse-field Ising model to the corresponding classical Ising model following this correspondence, spin-spin interactions arise in the additional dimension due to non-commutativity.
In other words, non-commutativity can effectively produce correlations within the dimension.
We might thus infer that the scalar curvature can reflect such interactions, too.

The 0D transverse-field Ising model mathematically shows a critical behavior at zero temperature in the following sense.
The longitudinal magnetic susceptibility at $h=0$ can be calculated as 
\begin{equation}
    \chi_{0\mathrm{D}}(h=0) = \left. \pdv[2]{\psi_{0\mathrm{D}}}{h} \right|_{h=0} = \dfrac{1}{\Gamma} \tanh x.
\end{equation}
In the limit of zero temperature the susceptibility becomes
\begin{equation}
    \chi_{0\mathrm{D}}(h=0) \to \dfrac{1}{\Gamma} .
\end{equation}
This yields a critical behavior at $\Gamma=0$.
Although the critical point itself is unphysical, it may be interesting to see the critical behavior because this critical point corresponds to that of the 1D classical Ising model through the correspondence.
We should note that if we take the limit $h, \Gamma\to0$ before the zero-temperature limit, such a singularity does not appear. Generally, a criticality arises only in the suitable order of limiting.

To see the relation between the scalar curvature and the quantum phase transition, we first take the zero-temperature limit $r \to \infty$ and obtain
\begin{equation}\label{scalar curvature of 0D quantum Ising: absolute zero}
    R_{0\mathrm{D}} \simeq \dfrac{\beta}{4r}e^{2r},
\end{equation}
from which we confirm the form \eqref{R behavior at T=0} and identify $C_{0\mathrm{D}}$ as
\begin{equation}
    C_{0\mathrm{D}} = \dfrac{1}{4\sqrt{h^2+\Gamma^2}}
\end{equation}
In the immediate vicinity of the critical point $h=\Gamma=0$, we first set $h=0$ and then considering $\Gamma\to 0$, this behaves as
\begin{equation}
    C_{0\mathrm{D}} \sim \frac{1}{\Gamma}, 
\end{equation}
hence we see a critical behavior with an exponent of 1.

Now we are ready to check the quantum-classical correspondence for this.
The scalar curvature for the 1D classical Ising model satisfies the relation \eqref{R behavior2}, and its critical exponent is 1 from $\alpha=1$ \cite{janyszekRiemannianGeometryThermodynamics1989}.
We find the same exponent here and confirm the correspondence \eqref{R behavior expected}.

We next consider not to take the zero-temperature limit before taking $h, \Gamma \to 0$ to see the behavior of $R_{0\mathrm{D}}$ at a fixed nonzero temperature.
We now return to the Massiue potential, rather than the negative free energy.
Just expanding the scalar curvature \eqref{scalar curvature of 0D quantum Ising} at $r=0$ produces
\begin{equation}\label{scalar curvature of 0D quantum Ising: small r}
    R_{0\mathrm{D}} = \dfrac{4}{9} r^2 + \order*{r^4}.
\end{equation}
Hence, at nonzero temperatures, approaching the critical point, $R_{0\mathrm{D}}$ converges to $0$ instead of diverging.
This is consistent with the fact that the 0D transverse-field Ising model does not mathematically show a critical behavior at nonzero temperatures.

The limit $r\to0$ can also be viewed as infinite temperature $T\to \infty$ with a fixed longitudinal field $h$ and transverse field $\Gamma$.
From this perspective, when we take $r\to0$, the system shows the classical nature because thermal fluctuation instead of quantum fluctuation becomes dominant.
Thus, $R_{0\mathrm{D}}$ should vanish in the limit, with which the equation \eqref{scalar curvature of 0D quantum Ising: small r} is consistent.

\section{The one-dimensional transverse-field Ising model \label{sec4}}

Let $\sigma_i$ denote a spin 1/2 at the $i$-th site.
Then the 1D transverse-field Ising model with an interaction strength $J>0$ and transverse field $\Gamma>0$ is given by the Hamiltonian
\begin{equation}
    \hat{H}_{1\mathrm{D}} = -J \sum_{i=1}^N \hat{\sigma}_i^z \hat{\sigma}_{i+1}^z - \Gamma \sum_{i=1}^N \hat{\sigma}_i^x
\end{equation}
or
\begin{equation}\label{1D quantum Ising}
    -\beta \hat{H}_{1\mathrm{D}} = \theta \sum_{i=1}^N \hat{\sigma}_i^z \hat{\sigma}_{i+1}^z +x \sum_{i=1}^N \hat{\sigma}_i^x ,
\end{equation}
where $\theta := \beta J \in (0,\infty)$ and $x := \beta \Gamma \in (0,\infty)$ are the canonical parameters.
Another candidate of reasonable parameterization is the effective strength of the transverse field $g:=\Gamma/J$ instead of $x$ to consider the quantum phase transition at zero temperature of the model at $g=1$
In the thermodynamic limit, the potential per site of this model has been obtained as\cite{pfeuty1970one,suzuki2012quantum}
\begin{equation}\label{potential of 1D quantum Ising}
    \psi_{1\mathrm{D}}(\theta,x) := \dfrac{1}{\beta N} \ln Z = \dfrac{1}{\beta} \int_{-\pi}^{\pi} \dfrac{\mathrm{d}k}{2\pi} \ln(2\cosh \sqrt{\theta^2+x^2+2\theta x\cos k}) .
\end{equation}

At sufficiently low temperature, using the saddle-point method for large $\beta$, we can expand this potential as
\begin{align}\label{potential of 1D quantum Ising2}
    \psi_{\mathrm{1D}}(\theta,x) 
    = \dfrac{1}{\beta} \qty(\dfrac{2}{\pi} (\theta+x) E\qty(\dfrac{4\theta x}{(\theta+x)^2}) + \varepsilon +  o(\varepsilon)) ,
\end{align}
where 
\begin{equation}
    \varepsilon := e^{-2|\theta-x|}
\end{equation}
is the small quantity associated with the low-temperature expansion, and the function $E$ is the complete elliptic integral of the second kind defined as
\begin{equation}\label{complete elliptic integral of the second kind}
    E(m) := \int_0^{\pi/2} \mathrm{d}k \sqrt{1-m \sin^2k}.
\end{equation}

Because of the symmetry $\theta \leftrightarrow x$ in the potential function \eqref{potential of 1D quantum Ising}, it is sufficient to consider the disorder case $J < \Gamma$ only.
To simplify expressions, we put
\begin{align}
    m(\theta,x) := \dfrac{4\theta x}{(\theta+x)^2} .
\end{align}
For the metric, we have
\begin{gather}
    \beta \qty(g^{\mathrm{q}}_{1\mathrm{D}})_{(0)\theta \theta} = \dfrac{(x^2+\theta^2)K(m(\theta,x))-(x+\theta)^2E(m(\theta,x))}{\pi \theta^2(x+\theta)}, \\
    \beta \qty(g^{\mathrm{q}}_{1\mathrm{D}})_{(0)\theta x} = -\dfrac{(x^2+\theta^2)K(m(\theta,x))-(x+\theta)^2E(m(\theta,x))}{\pi \theta x(x+\theta)}, \\
    \beta \qty(g^{\mathrm{q}}_{1\mathrm{D}})_{(0)x x} = \dfrac{(x^2+\theta^2)K(m(\theta,x))-(x+\theta)^2E(m(\theta,x))}{\pi x^2(x+\theta)}, \\
    \beta \qty(g^{\mathrm{c}}_{1\mathrm{D}})_{(1)\theta \theta} = -\beta \qty(g^{\mathrm{c}}_{1\mathrm{D}})_{(1)\theta x} = \beta \qty(g^{\mathrm{c}}_{1\mathrm{D}})_{(1)x x} = 4, 
\end{gather}
where $K(m)$ is the complete elliptic integral of the first kind defined as
\begin{equation}\label{complete elliptic integral of the first kind}
    K(m) := \int_0^{\pi/2} \dfrac{\mathrm{d}k}{\sqrt{1-m\sin^2k}} .
\end{equation}
Note that the formula for the derivatives of the complete elliptic integrals \eqref{complete elliptic integral of the second kind}, \eqref{complete elliptic integral of the first kind}
\begin{equation}\label{derivative of E}
    \dv{E(m)}{m} = \dfrac{E(m)-K(m)}{2m} ,
\end{equation}
\begin{equation}\label{derivative of K}
    \dv{K(m)}{m} = \dfrac{mK(m)-K(m)+E(m)}{2m(1-m)} 
\end{equation}
will be helpful, and that $\qty(g^{\mathrm{c}}_{1\mathrm{D}})_{(1)ij}$ can be obtained also by using \eqref{BKM metric: first order classical part} for the energy gap $2|J-\Gamma|$ of this model.
The leading term of the determinant of the metric \eqref{denominator: first order coefficient} becomes
\begin{equation}
    \det(g_{1\mathrm{D}}) \simeq \dfrac{4}{\beta^2} \dfrac{(x^2+\theta^2)K(m(\theta,x))-(x+\theta)^2E(m(\theta,x))}{\pi\theta^{2}x^{2}(x+\theta)} (x-\theta)^2 \varepsilon. \label{det. of BKM metric of 1D quantum Ising: 1st approx.}
\end{equation}

For the Christoffel symbols, we have
\begin{gather}
    \beta \qty(\psi^{\mathrm{q}}_{1\mathrm{D}})_{(0)\theta \theta \theta} = \dfrac{(2x^2-\theta^2)E(m(\theta,x))}{\pi \theta^3(x-\theta)} - \dfrac{(2x^2+\theta^2)K(m(\theta,x))}{\pi \theta^3(x+\theta)}, \\
    \beta \qty(\psi^{\mathrm{q}}_{1\mathrm{D}})_{(0)\theta \theta x} = \dfrac{xK(m(\theta,x))}{\pi \theta^2(\theta+x)} + \dfrac{xE(m(\theta,x))}{\pi \theta^2(\theta-x)}, \\
    \beta \qty(\psi^{\mathrm{q}}_{1\mathrm{D}})_{(0)\theta x x} = \dfrac{\theta K(m(\theta,x))}{\pi x^2(\theta+x)} + \dfrac{\theta E(m(\theta,x))}{\pi x^2(x-\theta)}, \\
    \beta \qty(\psi^{\mathrm{q}}_{1\mathrm{D}})_{(0)x x x} = \dfrac{(2\theta^2-x^2)E(m(\theta,x))}{\pi x^3(\theta-x)} - \dfrac{(2\theta^2+x^2)K(m(\theta,x))}{\pi x^3(x+\theta)}, \\
    \beta \qty(\psi^{\mathrm{c}}_{1\mathrm{D}})_{(1)\theta \theta \theta} = -\beta \qty(\psi^{\mathrm{c}}_{1\mathrm{D}})_{(1)\theta \theta x} = \beta \qty(\psi^{\mathrm{c}}_{1\mathrm{D}})_{(1)\theta x x} = -\beta \qty(\psi^{\mathrm{c}}_{1\mathrm{D}})_{(1)x x x} = 8.
\end{gather}

Therefore, using the equation \eqref{scalar curvature in 2-dim for Gibbs state} and considering sufficiently low temperature $\abs{\theta-x} \gg 1$, we obtain the leading term of the scalar curvature as
\begin{equation}\label{scalar curvature of 1D quantum Ising: absolute zero}
    R_{\mathrm{1D}} \simeq 
    \dfrac{1}{4\abs{J-\Gamma}}e^{2\abs{\theta-x}} = \dfrac{1}{4J\abs{1-g}}e^{2\theta\abs{1-g}}.
\end{equation}
Here we confirm the form \eqref{R behavior at T=0} for the scalar curvature and identify $C_{\mathrm{1D}}$ as
\begin{equation}
    C_{\mathrm{1D}} = \dfrac{1}{4J\abs{1-g}}
\end{equation}
Furthermore, in the immediate vicinity of the critical point $g= 1$, this quantity behaves as
\begin{equation}
    C_{\mathrm{1D}} \sim \dfrac{1}{\abs{1-g}},
\end{equation}
from which we observe a critical behavior and identify the critical exponent as 1

At first glance, the correspondence \eqref{R behavior expected} seems to break down with $\alpha=0$(log) for the corresponding 2D (square-lattice) classical Ising model.
However, the case of $\alpha \le 0$ requires attention because a discrepancy from the scaling form \eqref{R behavior2} is observed and $R \sim t^{\alpha-1}$ is shown instead for some models\cite{janke2002information,janke2003information}, including the 2D Ising model on a kagome lattice with $\alpha=0$\cite{mirza2013thermodynamic}.
Hence, it may be natural that we should expect the scalar curvature for the 1D transverse-field Ising model to admit $R \sim \gamma^{\alpha-1}$ rather than \eqref{R behavior expected} for the correspondence.
Our result indeed indicates this.

If we take the limit $\gamma \to 0$ without the zero-temperature limit, such a singularity is expected to vanish because of no quantum phase transition at nonzero temperatures.
A numerical evaluation supports this claim, although we will not discuss it in depth here.

Before we end this section, we study the high-temperature behavior of $R_{\mathrm{1D}}$.
The limit $T \to \infty$ physically means that the system should become classical and uncorrelated, thus $R_{\mathrm{1D}}$ expects to vanish, as well as $R_{\mathrm{0D}}$.
This can be confirmed directly as follows.
At high temperature $\theta, x \ll 1$, the Massieu potential $\beta \psi_{1\mathrm{D}}$ can be approximated as 
\begin{align}
    \beta \psi_{1\mathrm{D}}(\theta,x) 
    \simeq \ln 2 + \half(\theta^2+x^2) - \dfrac{1}{12}(\theta^4+4\theta^2x^2+x^4).
\end{align}
This shows that the metric is Euclidean up to the leading order, resulting in zero curvature.
Note that the leading term of $F$ at high temperature becomes
\begin{equation}
    F \simeq \dfrac{8}{9} (\theta^2+x^2),
\end{equation}
and hence, for $\theta, x \ll 1$,
\begin{equation}
    R_{1\mathrm{D}} \simeq \dfrac{4}{9} (\theta^2+x^2).
\end{equation}
Interestingly, we find exactly the same behavior, including its coefficient $\frac{4}{9}$, as $R_{0\mathrm{D}}$.

\section{Conclusion \label{sec5}}

In this paper, we have analytically calculated the scalar curvature induced from the BKM metric for the 0D and 1D transverse-field Ising models at low and high temperatures.
As a result, we observe that the scalar curvatures converge to zero in $T\to \infty$ and diverge exponentially in $T\to 0$ following \eqref{R behavior at T=0}.
This divergence will be attributed to quantumness, namely the non-commutativity of observables.
It should be pointed out that this behavior will be reminiscent of Balian's proposal of an interpretation of the scalar curvature of the BKM metric, which says that the scalar curvature quantitatively expresses how quantum a physical system is, and the larger it is, the more quantum the system will be\cite{balian2014entropy}.

Regarding the relation of the scalar curvature with the quantum phase transition, if we can redefine the criticality of the scalar curvature at $T=0$ as $C$, defined by \eqref{C}, then we observe critical behavior with an exponent of 1 for both models.
From this, we find that the 0D model is found to respect the correspondence \eqref{R behavior expected} straightforward, whereas the 1D model respects the modified scaling form $R \sim \abs{\Gamma-\Gamma_c}^{\alpha-1}$, which is a scaling form observed in thermodynamic geometry of some classical models with $\alpha \le 0$.

As the last remark, from the mathematical point of view, the Riemannian-geometric structure of the state space endowed with the BKM metric has been investigated, and the scalar curvature gains attention due to its monotone property\cite{petzGeometryCanonicalCorrelation1994,hiai1996curvature,dittmannCurvatureMonotoneMetrics2000}.
Such an approach might also be useful to capture the firm interpretation of the scalar curvature of the BKM metric for quantum statistical models.

\bmhead{Acknowledgements}


The author thanks Prof. Shogo Tanimura for reviewing the first manuscript and giving some comments.


\bmhead{Data Availability}
Data sharing is not applicable to this article as no data sets were generated or analyzed
during the current study.

\section*{Declarations}

\bmhead{Conflict of interest}
The author states that there is no conflict of
interest.




\begin{appendices}

\section{Derivation of \eqref{numerator: first order coefficient}}\label{secA1}





In this appendix, we will evaluate the asymptotic behavior of $F$, defined by \eqref{numerator}, at low temperature and derive \eqref{numerator: first order coefficient}.

In preparation for it, we first introduce the following useful relations obtained from the eigenvalue equation \eqref{eigenvalue equation}:
\begin{equation}\label{useful relation 1}
    \partial_i \mathcal{E}_n = (\mathcal{\hat{O}}_i)_{nn},
\end{equation}
\begin{equation}\label{useful relation 2}
    \braket{m}{\partial_i n} = \dfrac{(\mathcal{\hat{O}}_i)_{mn}}{\mathcal{E}_n-\mathcal{E}_m}.
\end{equation}
The relation \eqref{useful relation 1} is nothing but the so-called Hellmann-Feynman theorem.
The relation \eqref{useful relation 2} shows that parameter dependence of an energy eigenstate results in emerging non-diagonal elements in the energy basis and the factor $\mathcal{E}_n-\mathcal{E}_m$.
Note that using \eqref{useful relation 1}, we find
\begin{equation}\label{useful relation 3}
    \partial_i (\beta \Delta) = \partial_i (\mathcal{E}_0 - \mathcal{E}_1) = (\mathcal{\hat{O}}_i)_{00} - (\mathcal{\hat{O}}_i)_{11}
\end{equation}

Considering the discussion on the low-temperature behavior of the determinant of the BKM metric and looking at \eqref{denominator: first order coefficient}, we observe that the dominant part of the first-order coefficient of $\varepsilon$ at low temperature is the classical part, that is,
\begin{equation}\label{BKM metric: asymptotic behavior}
    g_{ij}(\theta) \simeq \dfrac{1}{\beta} \qty(\dfrac{1}{\beta} g_{(0)ij}^{\mathrm{q}}(\theta) + g_{(1)ij}^{\mathrm{c}}(\theta) \varepsilon),
\end{equation}
where the first-order coefficient can be, again, obtained as
\begin{equation}\label{BKM metric: first order classical part in Apnd}
    g_{(1)ij}^{\mathrm{c}} = (\partial_i \beta \Delta) (\partial_j \beta \Delta) = \qty((\mathcal{\hat{O}}_i)_{00} - (\mathcal{\hat{O}}_i)_{11}) \qty((\mathcal{\hat{O}}_j)_{00} - (\mathcal{\hat{O}}_j)_{11}).
\end{equation}
In other words, the coefficient of $\varepsilon$ consists only of diagonal elements of observables $\{\mathcal{\hat{O}}_i\}_i$ and does not involve non-diagonal elements.
Our objective here is to expand the same discussion for $\psi_{ijk}=\partial_k g_{ij}$.

From \eqref{BKM metric: asymptotic behavior}, we have
\begin{equation}
    \psi_{ijk}(\theta) = \partial_k g_{ij}(\theta) 
    \simeq \dfrac{1}{\beta^3} \psi_{(0)ijk}^{\mathrm{q}} 
    + \dfrac{1}{\beta} \qty{\partial_k \qty(g_{(1)ij}^{\mathrm{c}}(\theta)) + g_{(1)ij}^{\mathrm{c}}(\theta) \partial_k \qty(\beta \Delta)} \varepsilon
\end{equation}
where
\begin{equation}
    \dfrac{1}{\beta^3} \psi_{(0)ijk}^{\mathrm{q}} := \dfrac{1}{\beta} \partial_k \qty(\dfrac{1}{\beta} g_{(0)ij}^{\mathrm{q}}(\theta)) = \dfrac{1}{\beta} \partial_k \sum_{n>0} \dfrac{1}{\mathcal{E}_0 - \mathcal{E}_n} 2\Re (\mathcal{\hat{O}}_i)_{n0} (\mathcal{\hat{O}}_j)_{0n}.
\end{equation}
Thus, our task is to find the dominant term in the first-order coefficient of $\varepsilon$ in $\psi_{ijk}$.

From \eqref{useful relation 3} and \eqref{BKM metric: first order classical part in Apnd}, we immediately arrive at
\begin{equation}
    g_{(1)ij}^{\mathrm{c}}(\theta) \partial_k \qty(\beta \Delta) 
    = \qty((\mathcal{\hat{O}}_i)_{00} - (\mathcal{\hat{O}}_i)_{11}) \qty((\mathcal{\hat{O}}_j)_{00} - (\mathcal{\hat{O}}_j)_{11}) \qty((\mathcal{\hat{O}}_k)_{00} - (\mathcal{\hat{O}}_k)_{11}).
\end{equation}
This can be seen as being classical in that it apparently consists only of diagonal elements.
On the other hand, using \eqref{useful relation 2} and $0 = \partial_k \braket{n}{n} = \braket{\partial_k n}{n} + \braket{n}{\partial_k n}$, we have
\begin{align}
    \partial_k (\mathcal{\hat{O}}_i)_{nn} &= \partial_k \mel{n}{\mathcal{\hat{O}}_i}{n}
    = \mel{\partial_k n}{\mathcal{\hat{O}}_i}{n} + \mel{n}{\mathcal{\hat{O}}_i}{\partial_k n} \notag \\
    &= \sum_m \qty(\braket{\partial_k n}{m} \mel{m}{\mathcal{\hat{O}}_i}{n} + \mel{n}{\mathcal{\hat{O}}_i}{m} \braket{m}{\partial_k n}) \notag \\
    &= \sum_{m(\neq n)} \qty(\braket{\partial_k n}{m} \mel{m}{\mathcal{\hat{O}}_i}{n} + \mel{n}{\mathcal{\hat{O}}_i}{m} \braket{m}{\partial_k n}) \notag \\
    &= \sum_{m(\neq n)} \dfrac{(\mathcal{\hat{O}}_k)_{nm} (\mathcal{\hat{O}}_i)_{mn} + (\mathcal{\hat{O}}_i)_{nm} (\mathcal{\hat{O}}_k)_{mn}}{\mathcal{E}_n-\mathcal{E}_m},
\end{align}
from which we see that the derivative of a diagonal element involves non-diagonal elements only, and hence this is found to come purely from quantumness.
Therefore, at sufficiently low temperature $\beta \Delta \gg 1$, if $\partial_k ((\mathcal{\hat{O}}_i)_{00} - (\mathcal{\hat{O}}_i)_{11})$ asymptotically obeys $(\beta \Delta)^{a}$ with some $a>0$, we expect $\partial_k g_{(1)ij}^{\mathrm{c}} \ll g_{(1)ij}^{\mathrm{c}} \partial_k (\beta \Delta)$, and the dominant part in the first-order coefficient of $\varepsilon$ in $\psi_{ijk}$ becomes the classical term
\begin{align}
    \psi_{(1)ijk}^{\mathrm{c}} :=& g_{(1)ij}^{\mathrm{c}} \partial_k (\beta \Delta) \notag \\
    =& (\partial_i \beta \Delta) (\partial_j \beta \Delta) (\partial_k \beta \Delta) \notag \\
    =& \qty((\mathcal{\hat{O}}_i)_{00} - (\mathcal{\hat{O}}_i)_{11}) \qty((\mathcal{\hat{O}}_j)_{00} - (\mathcal{\hat{O}}_j)_{11}) \qty((\mathcal{\hat{O}}_k)_{00} - (\mathcal{\hat{O}}_k)_{11}).
\end{align}
This leads us to have
\begin{equation}
    \psi_{ijk}(\theta) 
    \simeq \dfrac{1}{\beta} \qty(\dfrac{1}{\beta^2} \psi_{(0)ijk}^{\mathrm{q}} 
    + \psi_{(1)ijk}^{\mathrm{c}} \varepsilon)
\end{equation}
effectively when we evaluate $F$.

We now move on to the evaluation.
Before that, we recall
\begin{equation}
    \lim_{T\to0} F = \mdet{g_{(0)11}^{\mathrm{q}} & g_{(0)12}^{\mathrm{q}} & g_{(0)22}^{\mathrm{q}} \\ \psi_{(0)111}^{\mathrm{q}} & \psi_{(0)112}^{\mathrm{q}} & \psi_{(0)122}^{\mathrm{q}} \\ \psi_{(0)112}^{\mathrm{q}} & \psi_{(0)122}^{\mathrm{q}} & \psi_{(0)222}^{\mathrm{q}} } = 0,
\end{equation}
which can be obtained from the constraint \eqref{constraint: psi at T=0}.
From the multilinearity of a determinant, we have
\begin{align}
    \frac{F}{\varepsilon}
    &\simeq 
    \frac{\mdet{\dfrac{1}{\beta} \qty(\dfrac{1}{\beta} g_{(0)11}^{\mathrm{q}} + g_{(1)11}^{\mathrm{c}}\varepsilon) %
    & \dfrac{1}{\beta} \qty(\dfrac{1}{\beta} g_{(0)12}^{\mathrm{q}} + g_{(1)12}^{\mathrm{c}}\varepsilon) %
    & \dfrac{1}{\beta} \qty(\dfrac{1}{\beta} g_{(0)22}^{\mathrm{q}} + g_{(1)22}^{\mathrm{c}}\varepsilon) \\%
    \dfrac{1}{\beta} \qty(\dfrac{1}{\beta^2} \psi_{(0)111}^{\mathrm{q}} 
    + \psi_{(1)111}^{\mathrm{c}} \varepsilon) & %
    \dfrac{1}{\beta} \qty(\dfrac{1}{\beta^2} \psi_{(0)112}^{\mathrm{q}} 
    + \psi_{(1)112}^{\mathrm{c}} \varepsilon) & %
    \dfrac{1}{\beta} \qty(\dfrac{1}{\beta^2} \psi_{(0)122}^{\mathrm{q}} 
    + \psi_{(1)122}^{\mathrm{c}} \varepsilon) \\ %
    \dfrac{1}{\beta} \qty(\dfrac{1}{\beta^2} \psi_{(0)112}^{\mathrm{q}} 
    + \psi_{(1)112}^{\mathrm{c}} \varepsilon) & %
    \dfrac{1}{\beta} \qty(\dfrac{1}{\beta^2} \psi_{(0)122}^{\mathrm{q}} 
    + \psi_{(1)122}^{\mathrm{c}} \varepsilon) & %
    \dfrac{1}{\beta} \qty(\dfrac{1}{\beta^2} \psi_{(0)222}^{\mathrm{q}} 
    + \psi_{(1)222}^{\mathrm{c}} \varepsilon) }}{\varepsilon} \notag \\[2ex]
    &\simeq 
    \dfrac{1}{\beta^3} \qty( %
    \mdet{\dfrac{1}{\beta}g_{(0)11}^{\mathrm{q}} & \dfrac{1}{\beta}g_{(0)12}^{\mathrm{q}} & \dfrac{1}{\beta}g_{(0)22}^{\mathrm{q}} \\ %
    \psi_{(1)111}^{\mathrm{c}} & \psi_{(1)112}^{\mathrm{c}} & \psi_{(1)122}^{\mathrm{c}} \\ %
    \dfrac{1}{\beta^2}\psi_{(0)112}^{\mathrm{q}} & \dfrac{1}{\beta^2}\psi_{(0)122}^{\mathrm{q}} & \dfrac{1}{\beta^2}\psi_{(0)222}^{\mathrm{q}} } %
    + \mdet{\dfrac{1}{\beta}g_{(0)11}^{\mathrm{q}} & \dfrac{1}{\beta}g_{(0)12}^{\mathrm{q}} & \dfrac{1}{\beta}g_{(0)22}^{\mathrm{q}} \\ %
    \dfrac{1}{\beta^2}\psi_{(0)111}^{\mathrm{q}} & \dfrac{1}{\beta^2}\psi_{(0)112}^{\mathrm{q}} & \dfrac{1}{\beta^2}\psi_{(0)122}^{\mathrm{q}} \\ %
    \psi_{(1)112}^{\mathrm{c}} & \psi_{(1)122}^{\mathrm{c}} & \psi_{(1)222}^{\mathrm{c}} } %
    ) \notag \\
    & \quad + \dfrac{1}{\beta^3} \mdet{g_{(1)11}^{\mathrm{c}} & g_{(1)12}^{\mathrm{c}} & g_{(1)22}^{\mathrm{c}} \\ %
    \dfrac{1}{\beta^2}\psi_{(0)111}^{\mathrm{q}} & \dfrac{1}{\beta^2}\psi_{(0)112}^{\mathrm{q}} & \dfrac{1}{\beta^2}\psi_{(0)122}^{\mathrm{q}} \\ %
    \dfrac{1}{\beta^2}\psi_{(0)112}^{\mathrm{q}} & \dfrac{1}{\beta^2}\psi_{(0)122}^{\mathrm{q}} & \dfrac{1}{\beta^2}\psi_{(0)222}^{\mathrm{q}} }.
\end{align}
If we can expect $\frac{1}{\beta}g_{(0)ij}^{\mathrm{q}} \cdot \psi_{(1)klm}^{\mathrm{c}} \gg g_{(1)ij}^{\mathrm{c}} \cdot \frac{1}{\beta^2}\psi_{(0)klm}^{\mathrm{q}}$ at sufficiently low temperature, we can omit the third term and finally arrive at \eqref{numerator: first order coefficient}.
As a remark, if the third term of this were not dropped, the scalar curvature for the 1D transverse-field Ising model would become
\begin{equation}
    R_{1\mathrm{D}} \simeq \beta \dfrac{2\abs{\theta-x}-1}{8\abs{\theta-x}^2} e^{2\abs{\theta-x}},
\end{equation}
where the additional term $-1$ appears in the numerator.
Hence, at low temperature $\beta\Delta = 2\abs{\theta-x} \gg 1$, dropping this additional term seems to be reasonable.

\end{appendices}


\bibliography{Reference}

\end{document}